\documentclass[twocolumn,tighten]{aastex63}

\usepackage{natbib}
\usepackage{multirow}
\usepackage{amsmath}
\usepackage{hyperref}
\hypersetup{citecolor=blue,urlcolor=blue}
\received{April 1, 2020}
\revised{June 2, 2020}
\accepted{June 3, 2020}
\submitjournal{ApJ}

\shorttitle{Circumnuclear molecular gas in quasars and star-forming galaxies}
\shortauthors{T. Izumi et al.}

\begin{document}
\title{Circumnuclear Molecular Gas in Low-redshift Quasars and Matched Star-forming Galaxies}

\correspondingauthor{Takuma Izumi}
\email{takuma.izumi@nao.ac.jp}

\author[0000-0002-0786-7307]{Takuma Izumi}
\altaffiliation{NAOJ Fellow}
\affil{National Astronomical Observatory of Japan, 2-21-1 Osawa, Mitaka, Tokyo 181-8588, Japan}
\affil{Department of Astronomical Science, The Graduate University for Advanced Studies, SOKENDAI, 2-21-1 Osawa, Mitaka, Tokyo 181-8588, Japan }

\author{John D. Silverman}
\affil{Kavli Institute for the Physics and Mathematics of the Universe (Kavli-IPMU, WPI), The
University of Tokyo Institutes for Advanced Study, The University of Tokyo, Kashiwa, Chiba 277-8583, Japan}
\affil{Department of Astronomy, School of Science, The University of Tokyo, 7-3-1 Hongo, Bunkyo-ku, Tokyo 113-0033, Japan}

\author{Knud Jahnke}
\affil{Max Planck Institut f\"{u}r Astronomie, K\"{o}nigstuhl 17, D-69117 Heidelberg, Germany} 

\author{Andreas Schulze}
\affil{National Astronomical Observatory of Japan, 2-21-1 Osawa, Mitaka, Tokyo 181-8588, Japan} 

\author{Renyue Cen}
\affil{Department of Astrophysical Sciences, Princeton University, Princeton, NJ 08544, USA} 

\author{Malte Schramm}
\affil{National Astronomical Observatory of Japan, 2-21-1 Osawa, Mitaka, Tokyo 181-8588, Japan} 

\author{Tohru Nagao}
\affil{Research Center for Space and Cosmic Evolution, Ehime University, Matsuyama, Ehime 790-8577, Japan} 

\author{Lutz Wisotzki}
\affil{Leibniz-Institut f\"{u}r Astrophysik Potsdam (AIP), An der Sternwarte 16, D-14482 Potsdam, Germany} 

\author{Wiphu Rujopakarn}
\affil{Department of Physics, Faculty of Science, Chulalongkorn University, 254 Phayathai Road, Pathumwan, Bangkok 10330, Thailand}
\affil{National Astronomical Research Institute of Thailand (Public Organization), Don Kaeo, Mae Rim, Chiang Mai 50180, Thailand}
\affil{Kavli Institute for the Physics and Mathematics of the Universe (Kavli-IPMU, WPI), The
University of Tokyo Institutes for Advanced Study, The University of Tokyo, Kashiwa, Chiba 277-8583, Japan}

\begin{abstract}
A series of gravitational instabilities in a circumnuclear gas disk (CND) 
are required to trigger gas transport to a central supermassive black hole (SMBH) and ignite Active Galactic Nuclei (AGNs). 
A test of this scenario is to investigate whether an enhanced molecular gas mass surface density 
($\Sigma_{\rm mol}$) is found in the CND-scale of quasars relative to a comparison sample of inactive galaxies. 
Here we performed sub-kpc resolution CO(2--1) observations with ALMA of four low-redshift ($z \sim 0.06$), 
luminous ($\sim 10^{45}$ erg s$^{-1}$) quasars 
with each matched to a different star-forming galaxy, having similar redshift, stellar mass, and star-formation rate. 
We detected CO(2--1) emission from all quasars, which show diverse morphologies. 
Contrary to expectations, $\Sigma_{\rm mol}$ of the quasar sample, 
computed from the CO(2--1) luminosity, tends to be smaller than the comparison sample at $r < 500$ pc; 
there is no systematic enhancement of $\Sigma_{\rm mol}$ in our quasars. 
We discuss four possible scenarios that would explain the lower molecular gas content 
(or CO(2--1) luminosity as an actual observable) at the CND-scale of quasars, 
i.e., AGN-driven outflows, gas-rich minor mergers, time-delay between 
the onsets of a starburst-phase and a quasar-phase, and X-ray-dominated region (XDR) effects on the gas chemical abundance and excitation. 
While not extensively discussed in the literature, XDR effects can have an impact on molecular mass measurements 
particularly in the vicinity of luminous quasar nuclei; therefore higher resolution molecular gas observations, which are now viable using ALMA, need to be considered. 
\end{abstract}
\keywords{galaxies: active --- galaxies: ISM --- galaxies: evolution --- quasars: general --- ISM: molecules}

\section{Introduction}\label{sec1}
How are supermassive black holes (SMBHs) in galaxies fed? 
This has been one of the key open questions in astrophysics 
ever since quasars and active galactic nuclei (AGNs) were firmly established 
to be powered by accretion onto SMBHs. 
Fueling material needs to lose $>$99\% of its angular momentum 
to travel from $\sim 10$ kpc-scale galactic disks all the way 
down to the central black hole \citep{1971MNRAS.152..461L}. 
Therefore, a specific physical mechanism is required to provide torques 
that can transport gas to the nuclear region where viscous forces in an accretion disk 
can then take over \citep[e.g.,][]{1998RvMP...70....1B}. 

In principle, major gas-rich galaxy mergers \citep[such as observed as ultra-luminous infrared galaxies = ULIRGs,][]{1996ARA&A..34..749S} could provide one such mechanism, 
generating torques leading to massive gas inflows \citep[e.g.,][]{1989Natur.340..687H,2005Natur.433..604D,2006ApJS..163....1H,2008ApJS..175..356H}. 
Recent high resolution and/or high sensitivity observations indeed 
show enhanced AGN fraction in major merger systems \citep{2011MNRAS.418.2043E,2011ApJ...743....2S,2018Natur.563..214K,2018PASJ...70S..37G}, 
which is particularly the case for dust-reddened quasars \citep[e.g.,][]{2008ApJ...674...80U}. 
While capable of triggering AGN, major mergers are not likely the {\it dominant} mechanism 
for fueling most SMBHs out to $z \sim 2$ \citep[e.g.,][]{2011ApJ...726...57C,2012MNRAS.425L..61S,2012ApJ...744..148K,2016ApJ...830..156M}. 
It appears that mergers account for about $\sim 20\%$ of all AGN activity \citep{2011ApJ...743....2S}. 

For nearby low-luminosity AGNs (i.e., Seyfert-class; nuclear bolometric luminosity $\lesssim 10^{44}$ erg s$^{-1}$), 
secular processes, induced by for example, 
a barred gravitational potential, galaxy--galaxy interaction 
\citep[e.g.,][]{1990Natur.345..679S,2004ARA&A..42..603K,2006ApJS..166....1H}, 
or minor mergers \citep[e.g.,][]{1994ApJ...425L..13M,1999ApJ...524...65T,2014MNRAS.440.2944K}, 
may be sufficient to redistribute gas in the galaxy and transport angular momentum outward. 
Recent multi-scale hydrodynamical simulations \citep[e.g.,][]{2010MNRAS.407.1529H} 
predict that such gravitational instabilities are also the mechanism 
for gas transport in luminous AGNs (i.e., quasar-class) that dominate black hole growth. 
The simulations show that when the gas mass surface density is sufficiently high 
in the central $\sim$ several $\times$ 100 pc region of a gaseous {\it circumnuclear disk} (CND), 
a series of instabilities occur, which could then reach gas down to sub-pc scales. 
In this picture, {\it black hole accretion generally takes place if sufficient gas 
is deposited at the CND-scale irrespective of the mechanism}. 

How much gas is deposited at the CND-scale and on what timescale 
in the hosts of luminous quasars are currently unanswered questions. 
These most likely further depend on redshift, mass, and local environment. 
In nearby Seyfert galaxies, \citet{2016ApJ...827...81I} found 
a positive correlation between dense molecular gas mass of CND and AGN luminosity, 
supporting the importance of circumnuclear gas amount in AGN fueling. 
For luminous quasars at $z \gtrsim 0.1$, 
molecular gas observations have been limited 
in resolution to $\gtrsim$ a few arcsec ($>$ kpc) thus far 
\citep{2003ApJ...585L.105S,2001AJ....121.1893E,2006AJ....132.2398E,2019arXiv191200085S}, 
probing only the total gas content. 

Now the Atacama Large Millimeter/submillimeter Array (ALMA) 
permits us to spatially resolve the central sub-kpc of such quasar-host galaxies. 
According to the simulations \citep[e.g.,][]{2010MNRAS.407.1529H}, 
there should be a threshold in the nuclear gas density, above which 
self-gravitating instabilities can form, and below which inflows essentially do not exist. 
As a first step to test this scenario, here we study whether significantly higher 
gas mass surface densities, which are the key parameter determining gravitational instability, 
are observed by ALMA at the CND-scale in luminous quasars 
(accretion rate $> 10\%$ of the Eddington-limited value) relative to comparison galaxies without AGN. 
Note that the actual accretion rate should fluctuate strongly 
\citep[e.g.,][]{2011ApJ...737...26N,2015MNRAS.451.2517S}: 
an observed intermediate accretion rate could either be due to a genuinely low inflow rate, 
or just to a temporally fluctuated low accretion rate for a very high inflow rate. 
This ambiguity can be reduced in a statistical sense when we observe more and more luminous quasars. 

This effort is organized as follows. 
In Section \ref{sec2}, we describe the details of our sample selection and ALMA observations. 
The observed properties are presented in Section \ref{sec3}. 
We discuss hypothesized differences in CND-scale gas mass surface density 
between quasars and comparison galaxies in Section \ref{sec4}. 
Our conclusions are summarized in Section \ref{sec5}. 
Throughout this paper, we adopt the concordant cosmological parameters 
$H_0$ = 70 km s$^{-1}$ Mpc$^{-1}$, $\Omega_{\rm M}$ = 0.3, and $\Omega_{\rm \Lambda}$ = 0.7.

\section{Data description}\label{sec2}
\subsection{Sample Selection}\label{sec2.1}
\begin{figure}
\begin{center}
\includegraphics[width=\linewidth]{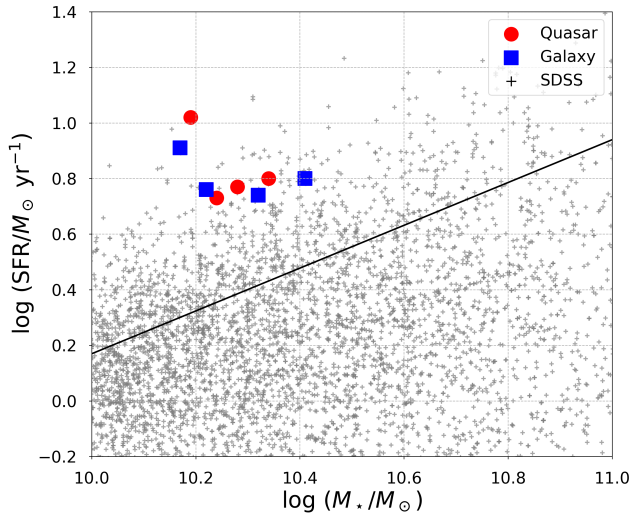}
\caption{
Location of our quasar and comparison galaxy samples in 
the stellar mass ($M_\star$) -- star-formation rate (SFR) plane. 
The red circles show the quasar sample while the blue squares show the comparison sample 
matched in $M_\star$, SFR, and $z$, respectively. 
The background gray points indicate the parent SDSS galaxies at $0.05 < z < 0.08$, 
which we used to draw the comparison sample. 
The black line indicates the $z \sim 0$ star-forming main sequence \citep{2007A&A...468...33E}. 
}
\label{fig1}
\end{center}
\end{figure}

To test the above hypothesis, we selected quasars 
and inactive galaxies, matched in redshift, stellar mass ($M_\star$), 
and global star formation rate (SFR) (Figure~\ref{fig1}; Table \ref{tbl1}). 
This leaves central gas surface density and AGN activity as prime parameters: 
a positive correlation between these therefore implies a causation. 
We initially selected $N = 5$ quasar-galaxy pairs for this purpose, 
but later realized that one pair is not well matched in SFR actually. 
Hence, we excluded that pair from our work, which leaves $N = 4$ pairs 
(see Appendix for details of the excluded pair). 
With this sample size, we can rule out the possibility 
that all quasars have higher gas surface density than galaxies solely by chance 
at a coincidence level of $\sim 94\%$ (random probability = 0.5$^N$ = 6\%). 

For the quasar sample, we selected nearby ($z < 0.08$), 
massive \citep[black hole mass $M_{\rm BH} > 10^{7.7}~M_\odot$ 
measured with the H$\beta$-based single epoch method,][]{2006ApJ...641..689V}, 
high-accretion type-1 objects from the Palomar-Green (PG) 
quasar sample \citep{1983ApJ...269..352S} and the Hamburg/ESO (HE) 
Survey quasar sample \citep{2000A&A...358...77W,2010A&A...516A..87S} with declination $\delta < 15^\circ$. 
Their quasar bolometric luminosities ($L_{\rm Bol}$) are $\sim 10^{45}$ erg s$^{-1}$, 
which roughly correspond to the knee of the quasar luminosity 
function at this redshift range \citep[e.g.,][]{2009ApJ...690...20S,2009A&A...507..781S} 
and allow us to better avoid the aforementioned degeneracy due to time-fluctuation. 
Their $M_\star$ are inferred from the local mass relation between SMBH and their host galaxy \citep{2013ARA&A..51..511K}. 
We used the far-infrared luminosity estimated from $L_{\rm Bol}$ \citep{2007ApJ...666..806N,2013A&A...560A..72R} 
to compute SFRs of our quasars based on the Kennicutt-Schmidt relation \citep[e.g.,][]{2012ARA&A..50..531K}. 
Note that these quasars were originally ultraviolet-selected, 
hence were not biased by the amount of gas and dust in their host galaxies. 

\begin{figure}
\begin{center}
\includegraphics[width=\linewidth]{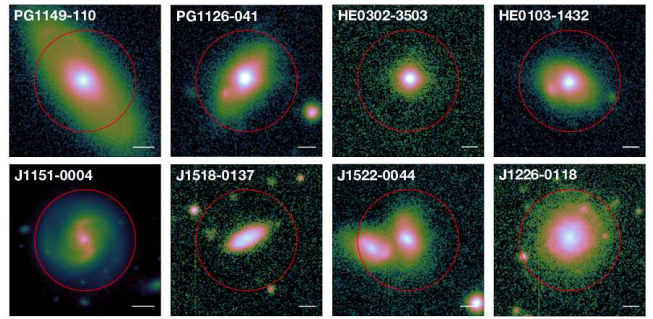}
\caption{
A gallery of our target quasars and inactive galaxies 
taken with the Subaru Hyper Suprime-Cam (J1151--0004; $R$-band) 
and ESO 2.2m Telescope WFI (the rest; ESO843 filter $\sim V$-band). 
Each panel shows $0.6\arcmin$ region centered on each nucleus, 
which are recorded in {\it GAIA} or SDSS database. 
The horizontal white scale bars correspond to 5 kpc length, 
and the red circles denote the nominal field of view of our ALMA observations ($\sim 24\arcsec$), respectively. 
}
\label{fig2}
\end{center}
\end{figure}

We constructed a comparison sample of inactive galaxies (in terms of quasar activity) 
selected from the SDSS DR10 MPA-JHU galaxy catalog \citep{2011AJ....142...72E,2004MNRAS.351.1151B}, 
which is matched to the quasar sample in fundamental parameters 
-- except SMBH accretion rate (Figure \ref{fig1}). 
Their $M_\star$ values are computed by SED fits to photometry following \citet{2003MNRAS.341...33K}. 
SFRs are measured with H$\alpha$ line emission for the SDSS galaxies \citep{2004MNRAS.351.1151B}. 
The constructed $N = 4$ pairs are exhibited in Figure \ref{fig2}. 

\begin{table*}
\begin{center}
\caption{Target quasars and comparison inactive galaxies \label{tbl1}}
\begin{tabular}{ccccccccccc}
\tableline\tableline
\multirow{2}{*}{Name} & R.A. & Decl. & \multirow{2}{*}{$z_{\rm opt}$} & Scale & \multirow{2}{*}{log ($\frac{M_{\rm BH}}{M_\odot}$)} & \multirow{2}{*}{ log ($\frac{L_{\rm Bol}}{\rm erg/s}$) } & \multirow{2}{*}{log ($\frac{M_\star}{M_\odot}$)} & \multirow{2}{*}{log ($\frac{SFR}{M_\odot /{\rm yr}}$)}\\ 
 & (ICRS) & (ICRS) &  & (kpc/$\arcsec$) & & & & & & \\ 
\tableline
PG1149$-$110 & 11:52:03.550 & $-$11:22:24.09 & 0.049 & 0.96 & 7.92 & 45.1 & 10.34 & 0.80 \\ 
PG1126$-$041 & 11:29:16.729 & $-$04:24:07.25 & 0.060 & 1.17 & 7.75 & 45.3 & 10.19 & 1.02 \\ 
HE0302$-$3503 & 03:04:26.924 & $-$34:52:07.66 & 0.066 & 1.27 & 7.86 & 45.0 & 10.28 & 0.77 \\ 
HE0103$-$1432 & 01:05:38.792 & $-$14:16:13.58 & 0.066 & 1.27 & 7.81 & 45.0 & 10.24 & 0.73 \\ \hline 
J1151$-$0004 & 11:51:30.954 & $-$00:04:39.93 & 0.048 & 0.93 & - & - & 10.41 & 0.80 \\ 
J1518$-$0137 & 15:18:34.689 & $-$01:37:43.83 & 0.063 & 1.20 & - & - & 10.17 & 0.91 \\ 
J1522$-$0044 & 15:22:24.740 & $-$00:44:04.58 & 0.067 & 1.28 & - & - & 10.32 & 0.74 \\ 
J1226$-$0118 & 12:26:46.972 & $-$01:18:54.13 & 0.062 & 1.20 & - & - & 10.22 & 0.76 \\ 
\tableline
\end{tabular}
\tablecomments{Top four objects are the target quasars and the bottom four are the comparison galaxies, 
which are listed in a paired order (e.g., PG1149$-$110 and J1151$-$0004). 
Their coordinates are tied to the {\it Gaia} reference system 
except for J1151$-$0004, J1522$-$0044, and J1226$-$0118 (SDSS system).}
\end{center}
\end{table*}

\subsection{ALMA Observations}\label{sec2.2}
We observed the redshifted $^{12}$CO(2--1) line (rest frequency $\nu_{\rm rest}$ = 230.5380 GHz) and its underlying continuum emission 
(at the rest frame wavelength $\lambda_{\rm rest} \sim$1.3 mm) towards our targets with ALMA, 
during Cycle 4 using the Band 6 receiver (project ID: \#2015.1.00872.S). 
As CO(2--1)/CO(1--0) line ratio (or excitation) is widely measured 
in various kinds of galaxies (see more details in \S~4), 
we can compute molecular masses by using this CO(2--1) line as a surrogate of the ground-transition CO(1--0) line, 
while easily acquiring higher angular resolutions than cases of lower frequency CO(1--0) observations. 
In total 37--43 antennas were used with the baseline 
ranging from 15.1 m to 1.1 km (0.7 km in some cases), 
resulting in a nominal maximum recoverable scale of $\sim 10\arcsec$. 
The reduction and calibration were done with CASA version 4.7 
\citep{2007ASPC..376..127M} in the standard manner. 
Continuum emission was subtracted in the $uv$-plane before making line cubes. 

All of the images presented in this paper were reconstructed 
using the task \verb|clean| with the natural weighting to enhance sensitivities, 
resulting in synthesized beam sizes of $\sim 0.34\arcsec$--$0.74\arcsec$ (major axes). 
The achieved 1$\sigma$ sensitivities are 0.26--0.41 mJy beam$^{-1}$ (quasars) 
and 0.16--0.56 mJy beam$^{-1}$ (comparison galaxies), respectively, for the CO(2--1) cubes: 
here we set the frequency resolution ($df$) to 62.5 MHz 
(velocity resolution $dV \simeq 86$ km s$^{-1}$) for quasars, 
whereas $df$ = 31.25 MHz ($dV \simeq 43$ km s$^{-1}$) for comparison galaxies, 
after finding that the lines are relatively fainter in the former (but with broader widths; see \S~3). 
Further details can be found in Table \ref{tbl2}. 
Note that on-source integration times were significantly shorter for the quasars (2--5 min) 
than for the comparison galaxies (4--12.5 min). 
The 10\% absolute flux uncertainty, according to the ALMA proposer's guide, 
is not included unless mentioned otherwise.

\section{Results}\label{sec3}
\subsection{CO(2--1) emission}\label{sec3.1}
\begin{figure*}
\begin{center}
\includegraphics[scale=0.85]{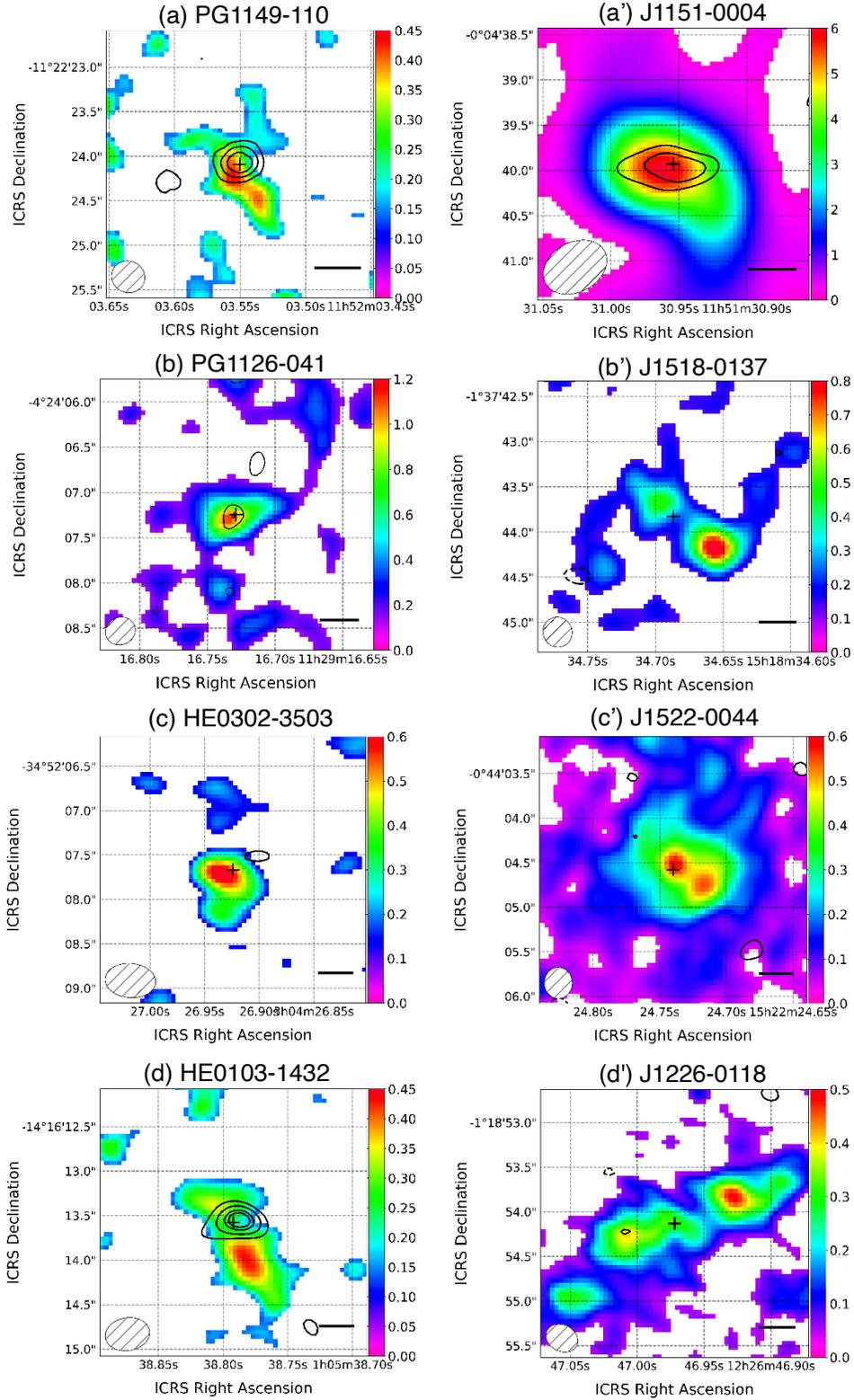}
\caption{
Velocity-integrated intensity maps of the CO(2--1) line emission (in units of Jy beam$^{-1}$ km s$^{-1}$) of 
the central 3$\arcsec$ region of our quasars (left column) and comparison galaxies (matched by letter; right column). 
The 1$\sigma$ sensitivity are (0.099, 0.092), (0.092, 0.039), (0.084, 0.023), and (0.090, 0.031) 
Jy beam$^{-1}$ km s$^{-1}$ for (a, a'), (b, b'), (c, c'), and (d, d'), respectively. 
The CO(2--1) maps start at 1.5$\sigma$ level to enhance the clarity. 
The central plus signs indicate the location of the quasar nuclei or galactic centers recorded in the {\it Gaia} or SDSS database. 
The contours indicate the continuum emission drawn at $-$3, 3, 4, 5, and 7$\sigma$ (see Table \ref{tbl2} for 1$\sigma$ values). 
The black horizontal bars correspond to 500 pc. 
}
\label{fig3}
\end{center}
\end{figure*}

\begin{figure}
\begin{center}
\includegraphics[scale=0.75]{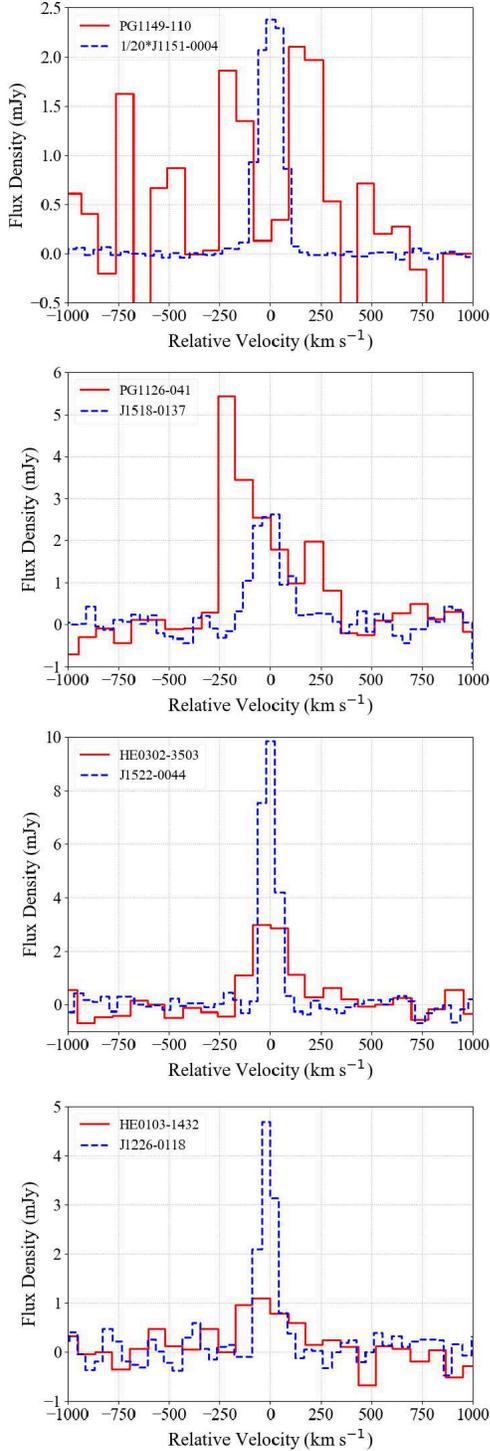}
\caption{
The CO(2--1) line spectra of the four pairs of quasars (red-solid line) and the comparison galaxies (blue-dashed line). 
The spectra in each panel were taken with the apertures matched to the poorer ones in each quasar-galaxy pair. 
Given the modest signal-to-noise ratio (e.g., PG1149$-$110) and asymmetric line profile (e.g., PG1126$-$041), 
we have not performed Gaussian fittings to these spectra. 
Note that the flux density of J1151$-$0004 is scaled by 1/20 to fit into the panel. 
}
\label{fig4}
\end{center}
\end{figure}

Figure \ref{fig3} shows the spatial distribution of the CO(2--1) line and underlying 1.3 mm continuum 
emission within the central 3$\arcsec$ region of our sample. 
Note that the coordinates of their centers are tied to the {\it Gaia} reference system 
except for J1151$-$0004, J1522$-$0044, and J1226$-$0118 (SDSS system). 
Their CO(2--1) line spectra are presented in Figure \ref{fig4}, 
which are used to determine the CO-based redshifts ($z_{\rm CO}$). 
The line emission was successfully detected in all of the targets 
and the line-emitting regions are spatially resolved 
in these global integrated intensity maps. 
We made these CO(2--1) maps (Figure \ref{fig3}) by integrating 
velocity channels over the line profiles. 
For the case of PG1149$-$110, in which the CO(2--1) line profile is not clear 
due to the modest signal-to-noise (S/N) ratio, 
we integrated over a velocity range of $\sim -400$ to $+400$ km s$^{-1}$. 

As for the quasar sample, line widths are broad with full-width at zero intensity (FWZI) $\sim 400-500$ km s$^{-1}$, 
which is consistent with the previous single dish-based CO observations 
toward optically-luminous PG/HE quasars \citep[$M_B < -20$ mag,][]{2007A&A...470..571B}. 
We also found that the line spectrum of PG1126$-$041, and likely that of 
PG1149$-$110, show double-horn profiles characteristic of a rotating CND. 
Recently, Atacama Compact Array (ACA) observations of CO(2--1) emission 
by \citet{2019arXiv191200085S} also reported a double-horn like profile 
for PG1126$-$041 but on a larger scale ($\sim 6\arcsec$ resolution). 

The CO line width is generally broader in the quasar sample than in the comparison sample. 
A line profile can be broader if the inclination angle of a system becomes higher, 
but we would not expect in terms of chance probability that all of our quasars 
have higher inclination angles than their matched galaxies. 
Hence, the broader line width would indicate higher gas rotation velocity 
and/or velocity dispersion in the quasar sample than in the comparison sample. 
One potential cause of a broader profile is a molecular outflow. 
However, it is hard to tell whether there are indications of outflows 
in our quasar spectra given the modest S/N ratios. 
Another possibility is that the quasars have larger enclosed mass (BH + stellar + gas) than comparison galaxies. 
In this case, the quasars may be in a later evolutionary phase than the star-forming comparison galaxies. 
A relevant discussion can be found in \S~4.

The morphology of the gas distribution is diverse with spiral-arm-like features (PG1126$-$041 and J1518$-$0137), 
bar- or edge-on disk-like structures (HE0103$-$1432 and J1226$-$0118), 
and face-on disk-like structures (HE0302$-$3503 and J1522$-$0044). In each case, we found that essentially 
all of the CO(2--1) emission we have recovered emerges from 
the central $\sim$a few kpc around the nuclei. 
Hereafter, we will focus on this small scale ($\lesssim 1\arcsec$) that is much smaller than our maximum 
recoverable scales (MRS), i.e., missing flux should not be an issue. 
This may start not to hold when we investigate larger scales, e.g., 
that is traced by the ACA \citep[aperture $\sim 6\arcsec$, see recent observations toward PG quasars in][]{2019arXiv191200085S}.
Hence we do not directly compare our results at this time with those of \citet{2019arXiv191200085S}. 

Following \citet{2005ARA&A..43..677S} we computed the CO(2--1) line luminosity ($L'_{\rm CO(2-1)}$) as 
\begin{equation}
\begin{split}
\left( \frac{L'_{\rm CO(2-1)}}{{\rm K~km~s^{-1}~pc^2}} \right) &= 3.25 \times 10^7~\left(\frac{S_{\rm CO(2-1)}\Delta V}{{\rm Jy~km~s^{-1}}} \right) \left( \frac{\nu_{\rm rest}}{{\rm GHz}} \right)^{-2} \\ 
& \left( \frac{D_L}{{\rm Mpc}} \right)^2 (1+z)^{-1},
\end{split}
\end{equation}
where $S_{\rm CO(2-1)}\Delta V$ is the integrated line flux and $D_L$ is the luminosity distance to the object. 
We first convolved the aperture of individual data cubes to the largest one 
among our full sample in physical scale, i.e., $\sim 700$ pc (J1151$-$0004). 
This sufficiently covers the typical CND-scale, as well as is roughly comparable to the spatial scale 
at which simulations \citep[e.g.,][]{2010MNRAS.407.1529H} start to see elevated gas surface density in quasars. 
The CO measurements within a central aperture for each object are summarized in Table \ref{tbl3}. 
CO line luminosities are $\sim (2-7) \times 10^7$ K km s$^{-1}$ pc$^2$ for the quasar sample 
and $\sim (3-19) \times 10^7$ K km s$^{-1}$ pc$^2$ for the comparison sample, respectively. 
Except for PG1126$-$041, all of our quasars show fainter $L'_{\rm CO(2-1)}$ 
than the matched galaxies (see \S~4 for further discussion). 
We stress that it is impossible to estimate how much of the total flux is concentrated at this small scale 
as there is no single dish CO(2--1) data available except for PG1126$-$041: 
in this particular case we found that $\sim 12\%$ of the total flux measured with the IRAM 30 m telescope 
\citep[14.9 Jy km s$^{-1},$][]{2007A&A...470..571B} originates from the central 700 pc.

\begin{table*}
\begin{center}
\caption{Descriptions of the CO(2--1) cubes and continuum maps \label{tbl2}}
\begin{tabular}{cccccc}
\tableline\tableline
\multirow{2}{*}{Name} & Beam sieze & Beam size & 1$\sigma$ (CO) & \multirow{2}{*}{$z_{\rm CO}$} & 1$\sigma$ (Cont.) \\ 
 & (CO: $\arcsec \times \arcsec$, $^\circ$) & (Cont.: $\arcsec \times \arcsec$, $^\circ$) & (mJy/beam) & & ($\mu$Jy/beam)  \\ 
\tableline
PG1149$-$110 & 0.38$\times$0.35, 69.5 & 0.34$\times$0.32, 73.3 & 0.41 & 0.0491 & 56.9 \\ 
PG1126$-$041 & 0.34$\times$0.30, $-$65.3 & 0.32$\times$0.29, $-$65.6 & 0.32 & 0.0603 & 47.3 \\ 
HE0302$-$3503 & 0.57$\times$0.39, 85.3 & 0.54$\times$0.37, 83.2 & 0.31 & 0.0656 & 38.7 \\ 
HE0103$-$1432 & 0.50$\times$0.37, $-$80.9 & 0.45$\times$0.34, $-$84.6 & 0.26 & 0.0669 & 42.8 \\ \hline
J1151$-$0004 & 0.74$\times$0.55, $-$61.0 & 0.72$\times$0.52, $-$59.5 & 0.56 & 0.0477 & 44.3 \\ 
J1518$-$0137 & 0.33$\times$0.32, 23.8 & 0.33$\times$0.31, 23.5 & 0.22 & 0.0628 & 26.2 \\ 
J1522$-$0044 & 0.35$\times$0.31, 0.1 & 0.33$\times$0.31, 2.9 & 0.16 & 0.0669 & 17.3 \\ 
J1226$-$0118 & 0.37$\times$0.30, 62.8 & 0.34$\times$0.29, 64.2 & 0.21 & 0.0625 & 22.5 \\ 
\tableline
\end{tabular}
\tablecomments{
1$\sigma$ sensitivities are determined in channels free of 
line emission extracted at the positions of the nuclei (CO cubes) 
and areas free of emission (continuum maps), respectively. 
The CO-based redshift ($z_{\rm CO}$) of each object was determined 
from an averaged frequency of channels containing $>2\sigma$ emission (see also Figure \ref{fig4}). 
}
\end{center}
\end{table*}

\begin{table*}
\begin{center}
\caption{CO line measurement and molecular gas mass \label{tbl3}}
\begin{tabular}{cccccccc}
\tableline\tableline
\multirow{3}{*}{Name} & $S_{\rm CO(2-1)}$ & $L'_{\rm CO(2-1)}$ & $M_{\rm mol}$ & $S_{\rm CO(2-1)}$ & $L'_{\rm CO(2-1)}$ & $M_{\rm mol}$ & $f_{0.7/2.0}$ \\ 
  & (Jy km s$^{-1}$) & (10$^7$ K km/s pc$^2$) & (10$^7$ $M_\odot$) & (Jy km s$^{-1}$) & (10$^7$ K km/s pc$^2$) & (10$^7$ $M_\odot$) & (\%)  \\ \cline{2-4} \cline{5-7}
  & \multicolumn{3}{c}{$\theta = 700$ pc} & \multicolumn{3}{c}{$\theta = 2$ kpc} & \\ 
\tableline
PG1149$-$110 & 0.74 $\pm$ 0.21 & 2.1 $\pm$ 0.6 & 2.7 & 1.66 $\pm$ 0.39 & 4.6 $\pm$ 1.1 & 6.0 & 45 $\pm$ 16  \\ 
PG1126$-$041 & 1.73 $\pm$ 0.21 & 7.3 $\pm$ 0.9 & 9.5 & 2.76 $\pm$ 0.44 & 11.6 $\pm$ 1.9 & 15.1  & 63 $\pm$ 13 \\ 
HE0302$-$3503 & 0.67 $\pm$ 0.12 & 3.4 $\pm$ 0.6 & 4.4 & 0.89 $\pm$ 0.15 & 4.5 $\pm$ 0.7 & 5.9  & 75 $\pm$ 18 \\ 
HE0103$-$1432 & 0.42 $\pm$ 0.10 & 2.2 $\pm$ 0.5 & 2.9 & 1.10 $\pm$ 0.23 & 5.7 $\pm$ 1.2 & 7.4 & 38 $\pm$ 12 \\ \hline 
J1151$-$0004 & 7.17 $\pm$ 0.72 & 18.7 $\pm$ 1.9 & 24.3 & 17.3 $\pm$ 0.2 & 45.2 $\pm$ 0.6 & 58.8 & 41 $\pm$ 4 \\ 
J1518$-$0137 & 0.87 $\pm$ 0.11 & 4.0 $\pm$ 0.5 & 5.2 & 3.96 $\pm$ 0.16 & 18.1 $\pm$ 0.7 & 23.5 & 22 $\pm$ 3 \\ 
J1522$-$0044 & 1.12 $\pm$ 0.11 & 5.8 $\pm$ 0.6 & 7.5 & 4.34 $\pm$ 0.14 & 22.6 $\pm$ 0.7 & 29.4 & 26 $\pm$ 3 \\ 
J1226$-$0118 & 0.66 $\pm$ 0.08 & 3.0 $\pm$ 0.4 & 3.9 & 2.66 $\pm$ 0.13 & 12.0 $\pm$ 0.6 & 15.6 & 25 $\pm$ 3 \\ 
\tableline
\end{tabular}
\tablecomments{
Apertures with which we measured the line luminosity are also indicated. 
The 10\% systematic flux uncertainty is included. 
The last column ($f_{0.7/2.0}$) shows the fraction of $L'_{\rm CO(2-1)}$ measured at the central 
700 pc relative to that measured at the central 2 kpc. 
}
\end{center}
\end{table*}

\subsection{Comments on J1151$-$0004}\label{sec3.2}
As shown in Figures \ref{fig3} or \ref{fig4}, one comparison galaxy J1151$-$0004 
is exceptionally bright in CO(2--1) emission ($L'_{\rm CO(2-1)} \simeq 2 \times 10^8$ K km s$^{-1}$ pc$^2$ at the central 700 pc). 
This galaxy was matched to the quasar PG1149$-$110, which shows much fainter CO(2--1) emission (Table \ref{tbl3}). 
The $L'_{\rm CO(2-1)}$ of J1151$-$0004 is very high even among the comparison sample as well. 

\begin{figure}
\begin{center}
\includegraphics[scale=0.3]{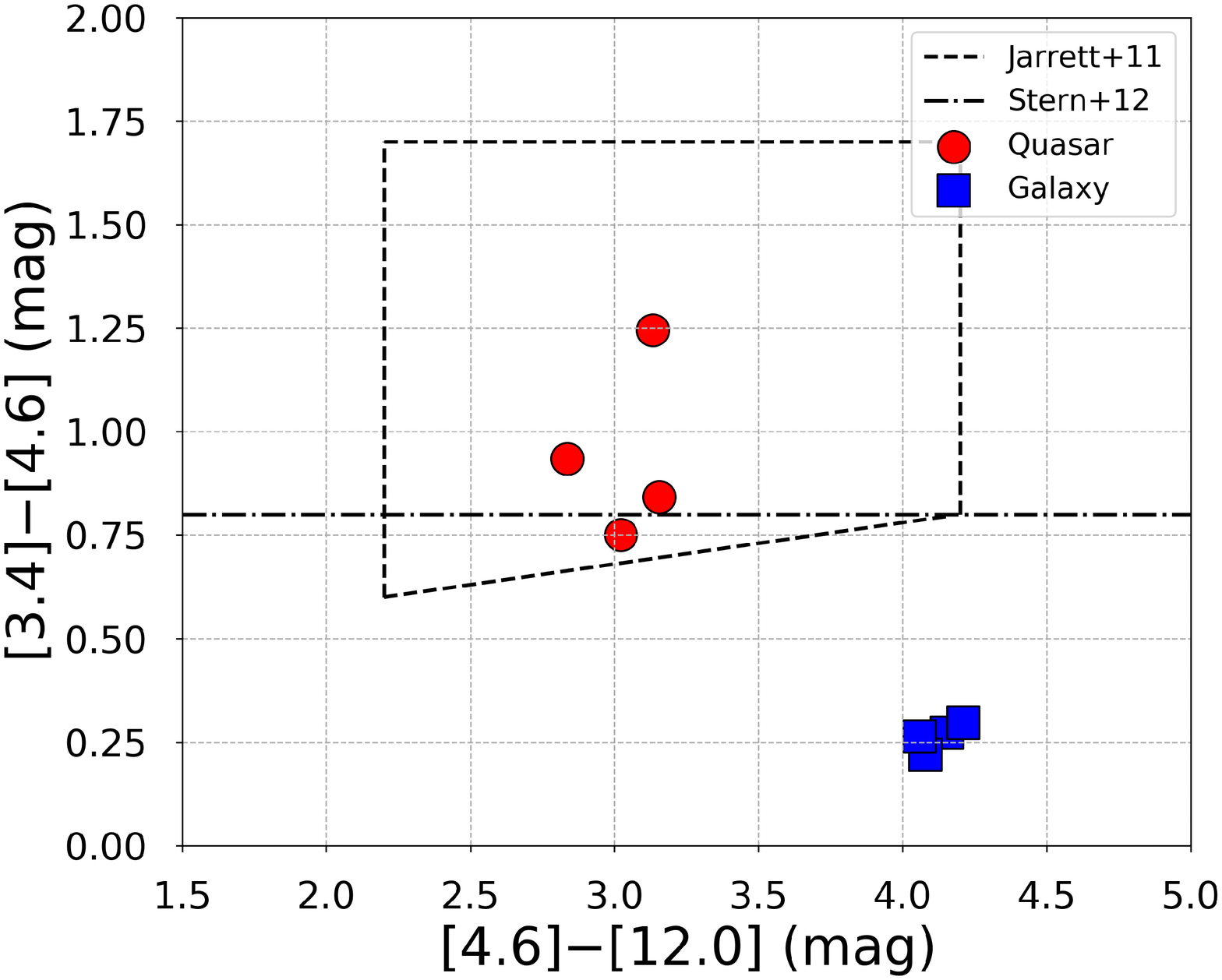}
\caption{
{\it WISE}-based 4.6 $-$ 12.0 $\mu$m vs. 3.4 $-$ 4.6 $\mu$m color-color diagram. 
Our target quasars (red circle) and comparison galaxies (blue square) are plotted. 
The black dashed and dot-dashed lines define the selection criteria of AGNs 
proposed by \citet{2011ApJ...735..112J} and \citet{2012ApJ...753...30S}, respectively. 
Our quasars (galaxies) satisfy the above selection criteria of being a quasar (galaxy). 
}
\label{fig5}
\end{center}
\end{figure}

For a possible explanation of this high $L'_{\rm CO(2-1)}$, we first considered 
that J1151$-$0004 hosts a significant level of dust-obscured star-formation, 
which is undetectable with SDSS. 
We further investigated the mid-infrared (MIR) photometric data of our comparison sample obtained by 
{\it Wide-field Infrared Survey Explorer} \citep[{\it WISE}:][]{2010AJ....140.1868W}, 
which is sensitive to dust-obscured activity. 
An estimated 8--1000 $\mu$m total IR luminosity ($L_{\rm TIR}$), using a tight correlation 
between $L_{\rm TIR}$ and {\it W3} band luminosity \citep{2017ApJ...850...68C}, 
is $L_{\rm TIR} = 7 \times 10^{10}~L_\odot$ ($W3$ magnitude = 8.35 mag) for the case of J1151$-$0004. 
This corresponds to SFR $\sim$ 10 $M_\odot$ yr$^{-1}$ \citep{2011ApJ...737...67M}, 
which is still comparable to the SDSS-based unobscured SFR (Table \ref{tbl1}). 
Moreover, the rest of the comparison galaxy sample also shows similar TIR-based SFR ($\sim 7-13~M_\odot$ yr$^{-1}$). 
Hence, the dust-obscured star-formation of J1151$-$0004 does not stand out among the comparison sample. 
Note that this measurement cannot be performed for the quasar sample 
as warm dust heated by the quasars themselves will dominate their {\it WISE} fluxes. 
We also considered the presence of a dust-obscured AGN as it may be hosted in a gas-rich galaxy. 
To briefly test this scenario, we measured {\it WISE} colors of our samples. 
We found that J1151$-$0004 has consistent {\it WISE} colors 
([4.6]$-$[12.0] vs [3.4]$-$[4.6] plane; Figure \ref{fig5}) with the rest of the galaxy sample, 
while our quasars show typical colors for quasars, indeed. 
Taking these into consideration, we suppose that J1151$-$0004 
is genuinely a star-forming galaxy that is exceptionally bright in CO(2--1). 
Any comparison with the matched quasar should be treated with care.

\subsection{Continuum emission}\label{sec3.3}
We detected $\lambda_{\rm rest} \simeq 1.3$ mm 
continuum emission significantly ($>$ 3$\sigma$) in PG1149$-$110, J1151$-$0004, and HE0103$-$1432, 
as well as marginally ($\sim 3\sigma$) in PG1126$-$041 (Figure \ref{fig3}). 
The nominal detection rate is 3 times higher for the quasar sample 
than for the comparison sample despite the shorter integration times for the former sample. 
The sizes of the continuum-emitting regions ($\lesssim 0\arcsec.5$) are smaller than the CO-emitting regions, i.e., more centrally concentrated. 
These likely imply a significant contribution of quasar-induced emission to the submillimeter continuum emission, 
either as additional heating to the thermal dust emission or as non-thermal synchrotron emission. 
Note that in the nearby luminous Seyfert galaxy NGC 1068, it is claimed 
that about half of ALMA Band 6 continuum flux measured around the nucleus 
is of non-thermal origin \citep{2014A&A...567A.125G}. 
As comparably high resolution radio-to-submillimeter continuum data is sparse for our targets, 
we do not perform detailed analysis to further reveal the nature of the Band 6 continuum emission. 

\subsection{Estimates of molecular gas mass}\label{sec3.4}
Molecular gas mass ($M_{\rm mol}$) is conventionally 
determined from CO(1--0) line luminosity $L'_{\rm CO(1-0)}$ 
by using a CO-to-molecular mass conversion factor $\alpha_{\rm CO}$ 
\citep[][and references therein]{2013ARA&A..51..207B}. 
Note that this molecular mass includes He and heavier elements in addition to H$_2$. 
To compute $M_{\rm mol}$, we first need to convert $L'_{\rm CO(2-1)}$ to $L'_{\rm CO(1-0)}$. 
We here assume $J$ = 2--1 to 1--0 CO brightness temperature ratio $R_{21} = 1$, 
which is the case of optically-thick and fully thermalized excitation. 
A value of $R_{21} \sim 1$ has been observed in the central regions of nearby 
star-forming galaxies \citep[e.g.,][]{2004A&A...427...45B,2013ApJ...777....5S}, 
IR-luminous galaxies \citep[e.g.,][]{2012MNRAS.426.2601P,2017ApJ...835..174S}, 
Seyfert galaxies \citep[e.g.,][]{2015ApJ...802...81M}, 
as well as in a global scale of high redshift submillimeter galaxies 
and luminous quasar host galaxies \citep{2013ARA&A..51..105C}. 
On the other hand, a lower $R_{21}$ of $\sim 0.5-0.8$ is frequently observed 
in $>$kpc scale galactic disks \citep[e.g.,][]{2013AJ....146...19L,2017ApJS..233...22S}. 
The latter value is also found for HE/PG quasars when observed at $>$a few kpc resolutions 
\citep{2007A&A...470..571B,2017MNRAS.470.1570H,2019arXiv191200085S}. 
Since we now investigate the central sub-kpc regions, however, we adopt $R_{21} = 1$ throughout this work. 

\citet{2013ApJ...777....5S} investigated 26 nearby star-forming galaxies 
and obtained 782 individual determinations of $\alpha_{\rm CO}$ coupled with a gas-to-dust ratio, 
after spatially-resolving the targets at some level. 
As an average for all of their values without weighting, 
$\alpha_{\rm CO} = 2.6~M_\odot$ pc$^{-2}$ (K km s$^{-1}$)$^{-1}$ 
is recommended, which has 0.4 dex standard deviation. 
This is somewhat lower than the Milky Way value of 
$4.3~M_\odot$ pc$^{-2}$ (K km s$^{-1}$)$^{-1}$ \citep{2013ARA&A..51..207B}, 
and is not strongly dependent on metallicity of galaxies as long as 
that metallicity is comparable to, or higher than, the solar value. 
In the central $\lesssim$kpc regions of galaxies, \citet{2013ApJ...777....5S} also found that 
$\alpha_{\rm CO}$ decreases by a factor of $\sim 2$ from the value averaged over the galaxies. 
Hence, we adopt $\alpha_{\rm CO} = 1.3~M_\odot$ pc$^{-2}$ (K km s$^{-1}$)$^{-1}$ in this work, 
which would have $\sim 0.4$ dex uncertainty as well. 
Note that we apply this $\alpha_{\rm CO}$ to both the quasar and the comparison galaxy samples, 
although AGN activity can potentially affect molecular gas properties 
including gas excitation and chemistry \citep[e.g.,][]{2013PASJ...65..100I,2016ApJ...818...42I}. 
As a consequence, we obtained $M_{\rm mol} = (2.7-9.5) \times 10^7~M_\odot$ for the quasar sample 
and $(3.9-24.3) \times 10^7~M_\odot$ for the galaxy sample, respectively, in the central 700 pc aperture (Table \ref{tbl3}).

\subsection{Molecular gas mass surface density profiles and circumnuclear gas}\label{sec4.1}
\begin{figure*}
\begin{center}
\includegraphics[width=\linewidth]{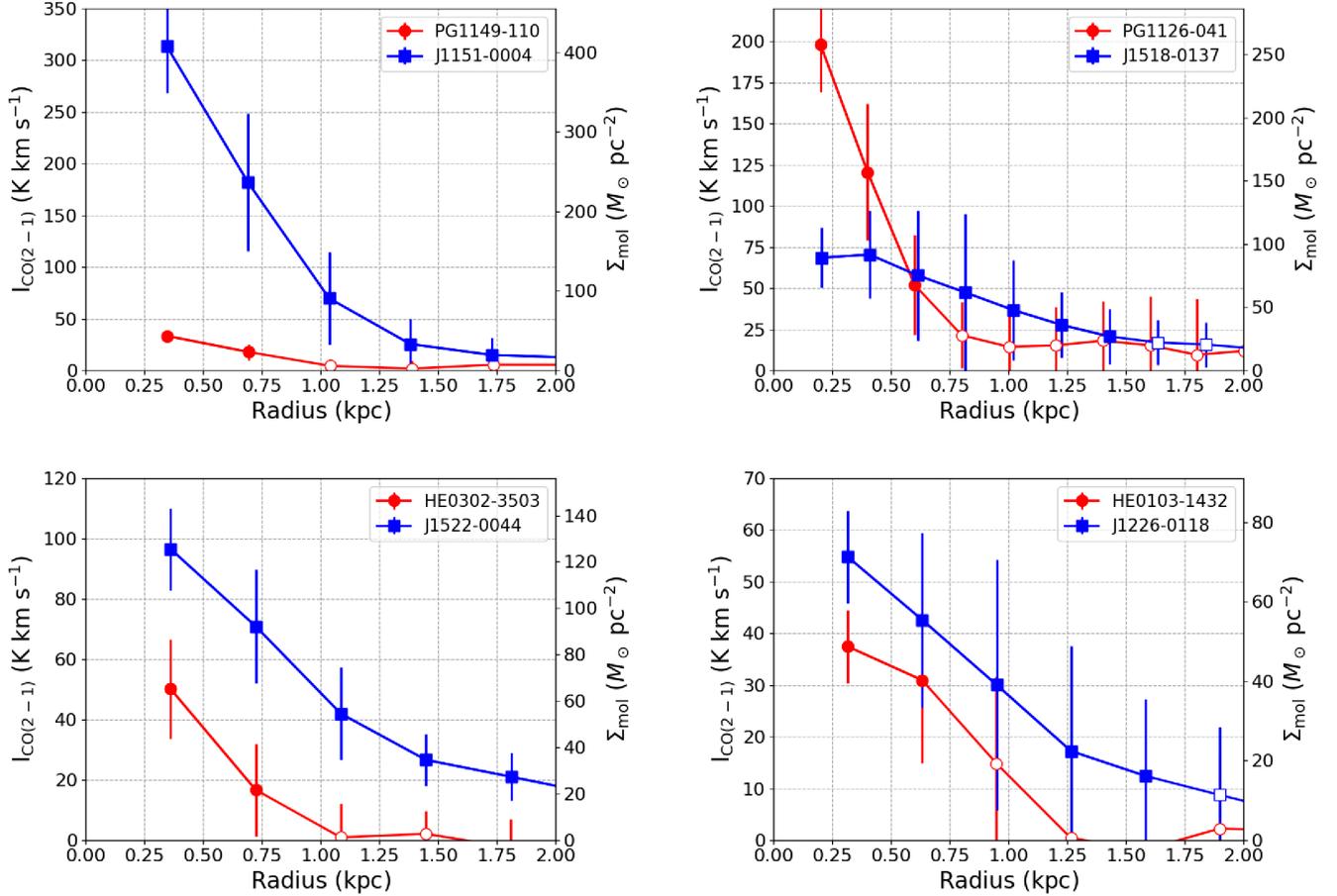}
\caption{
Azimuthally-averaged radial distributions of CO(2--1) integrated intensity 
and corresponding total molecular gas mass surface density of the matched pairs 
of quasars (red circles) and galaxies (blue squares) measured from each nucleus. 
Error bars indicate the scatter in each annulus. 
Open symbols indicate that the values are below 2$\sigma$ of our integrated intensity maps. 
Note that these are projected profiles on the sky without inclination corrections 
and the measurements were performed after matching the beam sizes in physical scale. 
}
\label{fig6}
\end{center}
\end{figure*}

A primary goal of this work is to test whether systematically enhanced circumnuclear 
gas mass surface densities are found in the quasar sample relative to the comparison sample. 
Figure \ref{fig6} shows the azimuthally-averaged radial distributions of 
CO(2--1) integrated intensity\footnote{We here describe this quantity in the brightness temperature unit $I_{\rm CO(2-1)}$ (K km s$^{-1}$) for an easy conversion to $M_{\rm mol}$.} 
and corresponding total molecular gas mass surface density ($\Sigma_{\rm mol}$) 
of the four matched pairs measured at the central $r \leq 2$ kpc. We used MIRIAD \citep{1995ASPC...77..433S} task \verb|ellint| to make these plots 
after matching the physical resolution (i.e., not angular resolution) to the poorer one of each pair. 
The steps between concentric rings correspond to $\theta_{\rm maj}/2$, 
where $\theta_{\rm maj}$ is the major axis of each 2D Gaussian beam. 
Note that these profiles are sky projections as we have no robust 
information about the inclination angles of our targets. 
However, the inclination angle (projection) effect will impact the final values of $\Sigma_{\rm mol}$. 
For example, if we assume that the classical AGN torus scheme 
\citep{1993ARA&A..31..473A} is applicable to our quasars, 
their inclination angle ($i$) may be $\lesssim 45\arcdeg-60\arcdeg$ 
\citep[characteristic value for type-1 AGNs,][]{2014MNRAS.441..551M}. 
In this case, $\Sigma_{\rm mol}$ could be further reduced by a factor $\sim 2$ (correction factor = cos $i$). 
Future high resolution observations that accurately constrain the gas distribution and/or dynamics 
are necessary to properly correct this effect both for the quasars and the inactive galaxies. 

From Figure \ref{fig6}, the quasar sample shows 
lower $\Sigma_{\rm mol}$ than the comparison sample in three of the four pairs: 
only PG1126$-$041 shows higher $\Sigma_{\rm mol}$ 
relative to its comparison galaxy J1518$-$0137 in the central region. 
This result would be robust even if we regard the galaxy J1151$-$0004 
to be an anomalous case based on its high $L'_{\rm CO(2-1)}$, 
as $\Sigma_{\rm mol}$ of the paired quasar PG1149$-$110 is still smaller 
than the remaining three comparison galaxies at the central $r \lesssim 500$ pc. 
Therefore, we do not find {\it systematically enhanced} 
$\Sigma_{\rm mol}$ in these quasars as compared to the comparison galaxies. 
This appears to be contradictory to our initial expectation 
that the gravitational instability at the CND-scale caused by rich amount of gas triggers AGN, 
which is predicted in various galaxy evolution models \citep[e.g.,][]{2010MNRAS.407.1529H}, 
as well as is likely supported by recent observations toward nearby Seyfert galaxies \citep{2016ApJ...827...81I}.

Regarding the comparison galaxy sample, their $\Sigma_{\rm mol}$ 
at the central $r \lesssim 500$ pc are larger than the typical values 
($\sim 50-100~M_\odot$ pc$^{-2}$; measured with the Galactic $\alpha_{\rm CO}$) found in circumnuclear regions 
of nearby star-forming galaxies \citep[e.g.,][]{2008AJ....136.2782L}. 
On the other hand, $\Sigma_{\rm mol}$ of our quasars, except for PG1126$-$041, 
tend to be comparable ($\lesssim 70~M_\odot$ pc$^{-2}$) to that typical value for star-forming galaxies 
\footnote{ 
Note that this is in stark contrast to some previous CO observations toward PG quasars 
\citep[e.g.,][]{2001AJ....121.1893E,2006AJ....132.2398E}, 
primarily as those quasars were selected based on their IR brightness.}, 
although the absolute value of $\Sigma_{\rm mol}$ 
critically depends on the adopted $\alpha_{\rm CO}$: 
the value can increase by, for example, $\sim 3\times$ when we adopt the Galactic $\alpha_{\rm CO}$. 
Indeed, recent hydrodynamic simulations predict that $\alpha_{\rm CO}$ has a large dispersion \citep[e.g.,][]{2018ApJ...852...88W}. 
As there is no effective way to estimate $\alpha_{\rm CO}$ in our targets at this moment, 
we use the currently adopted value in this work. 
Future multi-$J$ CO observations can lessen this source of uncertainty. 

The lower gas masses in these quasars are also evident on a kpc scale as the CO(2--1) emission 
of the comparison galaxy sample is brighter 
and spatially more extended than the quasar sample (Figure \ref{fig3}), 
although the sensitivity of the line cube is considerably different between the two samples (Table \ref{tbl2}). 
For a practical purpose, we measured $L'_{\rm CO(2-1)}$ and $M_{\rm mol}$ 
of the two samples using a common aperture of $\theta = 2$ kpc (Table \ref{tbl3}). 
The quasar sample shows $L'_{\rm CO(2-1)} = (4.5-11.6) \times 10^7$ K km s$^{-1}$ pc$^2$ 
(or $M_{\rm mol} = (6-15) \times 10^7~M_\odot$)\footnote{While we probe $>$kpc scales, we continue to use $\alpha_{\rm CO} = 1.3~M_\odot$ pc$^{-2}$ (K km s$^{-1}$)$^{-1}$ for simplicity in this work.}, 
which is clearly smaller than those of the comparison sample 
$L'_{\rm CO(2-1)} = (12.0-45.2) \times 10^7$ K km s$^{-1}$ pc$^2$ 
(or $M_{\rm mol} = (16-59) \times 10^7~M_\odot$). 
A similar trend was also found over global scales 
in $z \sim 1.5$ luminous (log ($L_{\rm Bol}$/erg s$^{-1}$) $>$ 45--46) quasars \citep{2017MNRAS.468.4205K}, 
and bulge-dominated HE quasars \citep{2017MNRAS.470.1570H}.

\section{Discussion}\label{sec4}
We consider four plausible scenarios to explain the molecular gas properties in the central regions 
of quasar hosts as compared to a matched control sample of star-forming galaxies. 
All cases need to consider the lower gas content, the dissimilar radial surface brightness profiles and different velocity profiles. 

\begin{figure}
\begin{center}
\includegraphics[scale=0.5]{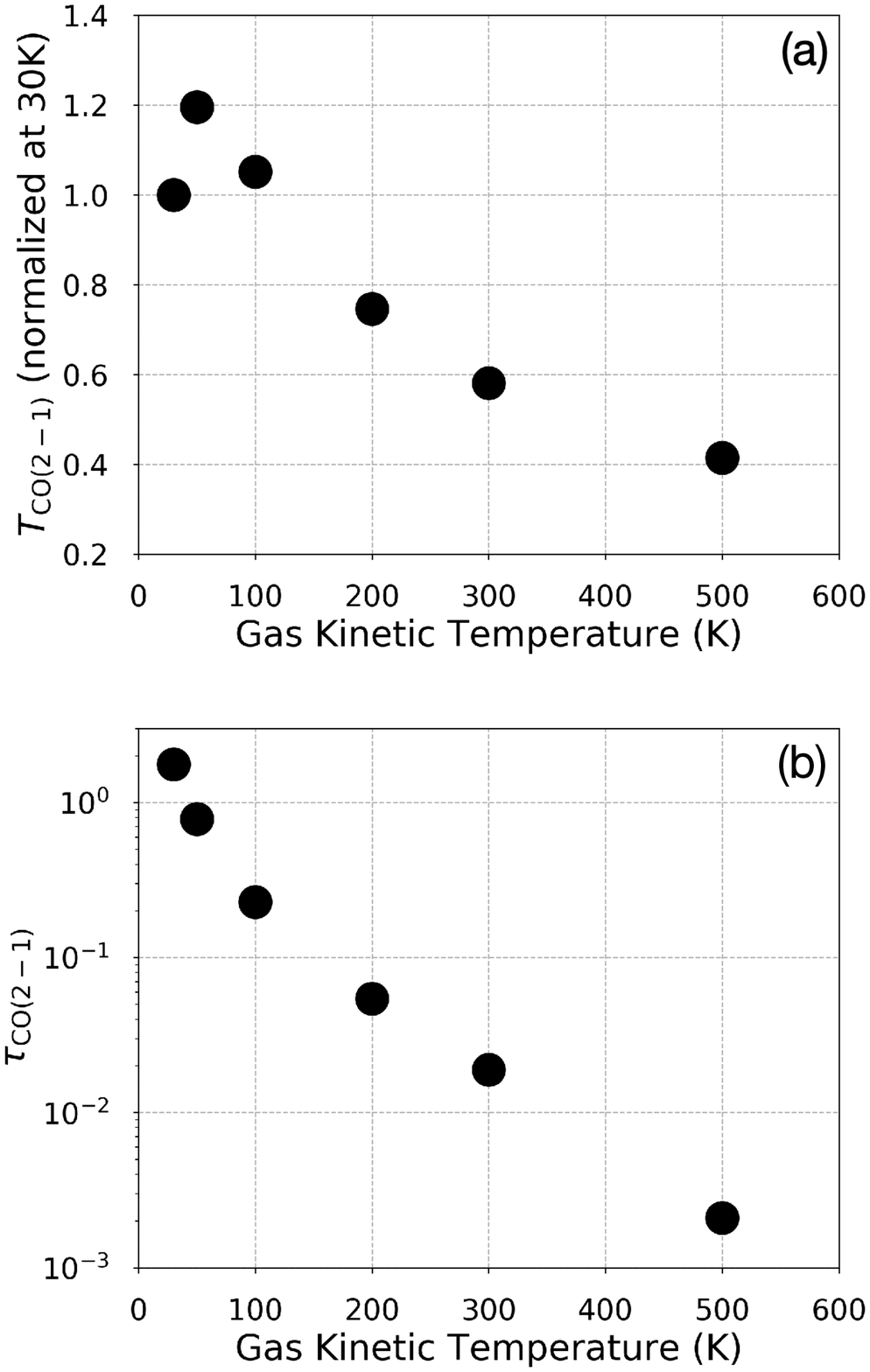}
\caption{
Non-LTE radiative transfer modelings of (a) CO(2--1) brightness temperature and (b) its line opacity ($\tau_{\rm CO(2-1)}$), 
as a function of gas kinetic temperature ($T_{\rm kin}$), performed with the RADEX code. 
The CO(2--1) brightness is normalized by the value at $T_{\rm kin} = 30$ K. 
For simplicity, we only show the results with H$_2$ gas density ($n_{\rm H2}$) of 10$^5$ cm$^{-3}$ and 
CO column density-to-velocity width ratio ($N_{\rm CO}/dV$) of $2.5 \times 10^{16}$ cm$^{-2}$ (km s$^{-1}$)$^{-1}$. 
}
\label{fig7}
\end{center}
\end{figure}

(i) {\it AGN-driven outflows --} massive outflows in various phases 
of gas have been observed both in nearby AGNs \citep[e.g.,][]{2012A&A...537A..44A,2012ApJ...746...86G,2014A&A...562A..21C} 
and in high redshift quasars \citep[e.g.,][]{2008A&A...491..407N,2012MNRAS.425L..66M,2019A&A...630A..59B}. 
Among the multiphase flows (ionized, atomic, and molecular), 
molecular outflows carry the bulk of the gas masses \citep[e.g.,][]{2019ApJ...871..156M}, 
which are considered to be significant enough to deplete CNDs. 
If we adopt the positive correlation between AGN luminosity and molecular outflow rate ($\dot{M}_{\rm H2}$) 
derived by \citet{2014A&A...562A..21C}, we expect as high as $\dot{M}_{\rm H2} \simeq 400~M_\odot$ yr$^{-1}$ for our quasar sample. 
Hence the gap of $M_{\rm mol}$ between the quasar sample and the comparison sample 
can be easily reconciled if such molecular outflows have lasted only for $\sim 1$ Myr, 
which is a small portion of a typical life-time of quasars \citep[$\sim 10-100$ Myr,][]{2017ApJ...847...81S}. 
We recall that we cannot discern the existence of such outflows with current data 
given the modest S/N ratios (Figure \ref{fig4}). 

A caveat of this scenario is that a large fraction of the CO(2--1) emission 
of our quasar sample comes from the very central region. 
For example, the ratio of $L'_{\rm CO(2-1)}$ measured 
with $\theta = 700$ pc to that measured with $\theta = 2$ kpc, 
which is denoted as $f_{\rm 0.7/2.0}$ in Table \ref{tbl3}, 
is systematically higher in the quasar sample ($\sim 38-75\%$; average = 52.6 $\pm$ 7.1\%) 
than in the comparison sample ($\sim 22-41\%$; average = 27.0 $\pm$ 1.6\%). 
We may expect an opposite case, i.e., lower central gas concentration in the quasar sample, 
as AGN-driven outflows basically expel their surrounding gas from inside to outside \citep{2015ARA&A..53..115K}. 
A recent hydrodynamic simulation also suggests that 
AGNs cause little impact on the surrounding material via winds \citep{2014MNRAS.441.1615G}. 
However, there may be a possibility that the cavity caused by the feedback 
is much smaller \citep[e.g., $<100$ pc;][]{2016MNRAS.458..816H} than our resolutions. 
Higher resolution observations are definitely needed. 

(ii) {\it Gas-rich minor merger --} 
the $M_{\rm mol}$ we find at the central 700 pc 
of the quasar sample are of the order of $10^7~M_\odot$. 
An average (bulge-scale) $M_\star$, expected for the quasar sample, 
is $\sim 2 \times 10^{10}~M_\odot$ (Table \ref{tbl1}). 
Hence, a 10:1 minor merger (i.e., $M_\star = 2 \times 10^9~M_\odot$) 
can provide that amount of molecular gas if we assume a typical gas mass fraction 
of $\sim$a few to $\sim 10$ \% for galaxies with $M_\star = 10^9 - 10^{10}~M_\odot$ \citep[e.g.,][]{2015MNRAS.454.3792M}. 
This scenario is in line with recent observational evidence that major mergers 
are not the dominant driver of quasar activity at least out to $z \sim 2$ 
\citep[e.g.,][]{2011ApJ...726...57C,2012MNRAS.425L..61S,2012ApJ...744..148K,2016ApJ...830..156M}. 
It is particularly noteworthy that if the merging (satellite) galaxy also hosts an SMBH,
it can form a binary system with the primary SMBH, 
which then causes gravitational instability in the newly formed gas disk \citep{1996ApJ...469..581T}. 
Very deep optical imaging observations that can capture evidence of past minor mergers are worth performing, 
such as done for the nearby Seyfert galaxy NGC 1068 \citep{2017PASJ...69...90T}. 

(iii) {\it Time-delay --}
Another possible explanation is that a quasar-phase happens during a longer time-scale starburst-phase \citep[e.g.,][]{2012MNRAS.420L...8H}. 
If this is the case, a large portion of the H$_2$ gas was already consumed by the preceding starburst event. 
Such time-delay has been predicted not only in galaxy-evolution models including merger-induced ones \citep[e.g.,][]{2005Natur.433..604D,2008ApJS..175..356H}, 
but also by actual observations \citep[e.g.,][]{2007ApJ...671.1388D,2016A&A...587A..72B}. 
Observations suggest that the time-difference between the onsets of the two phases is 
$\sim 250-500$ Myr \citep{2010MNRAS.405..933W,2010ApJ...714L.108S}. 
Hence, our quasar sample could have had another $\sim 10^9~M_\odot$ H$_2$ gas in the past, 
if the currently estimated SFRs (Table \ref{tbl1}) have lasted for a 100 Myr. 
This value can be even larger if the quasars have experienced significant 
reduction of SFR over their lives. 
In these cases, we expect significantly greater $\Sigma_{\rm mol}$ in the CNDs 
of our quasars at the time they were actually activated, 
than the currently observed modest values of $\Sigma_{\rm mol}$. 

This time-delay would thus wash out a causality between the CND-scale 
gas properties including $\Sigma_{\rm mol}$ and currently observed quasar activity. 
Hence, we need to perform a test, like what we did in this work, to AGNs in a much younger phase 
to better investigate the importance of CND-scale $\Sigma_{\rm mol}$ for triggering AGN events. 
Deeply buried (i.e., dust-obscured) AGNs with high Eddington ratios 
\citep[e.g.,][]{2007ApJS..171...72I,2004A&A...420L..23K} would be useful for this test. 
Note that there is a suggestion that the amount of gas reaching the nuclear and accretion disk 
would depend on the initial gas density profile of a galaxy \citep{2015ApJ...805L...9C}. 
The higher $f_{0.7/2.0}$ in the quasar sample than the comparison sample discovered in this work 
may still imply the initially-different density profiles we expect. 

One thing that needs to be further addressed is the relationship 
to what has been observed in Seyfert-class objects: 
at the CND-scale of Seyfert galaxies there is a tight positive correlation 
between SFR and mass accretion rate onto the AGN \citep[e.g.,][]{2012ApJ...746..168D,2014ApJ...780...86E}, 
or similarly dense molecular gas mass and accretion rate \citep{2016ApJ...827...81I}, 
while we do not see systematic enhancement in gas amount (or CO(2--1) luminosity as an actual observable here) 
between the quasar sample and even the non-AGN comparison sample. 
Regarding this, we speculate that the much higher AGN luminosity in quasars than Seyfert galaxies 
would explain this difference at some level as discussed in the following.

(iv) {\it XDR effects} -- Owing to the high X-ray flux expected around an AGN, 
there may form an X-ray dominated region (XDR), in which gas physical and chemical properties 
are governed by the X-ray irradiation \citep{1996ApJ...466..561M,2005A&A...436..397M}. 
While the actual size of the XDR depends on the incident X-ray radiation and gas density, 
typical values are expected to be several hundred pc (i.e., CND-scale we probed in this work) 
according to a theoretical work \citep{2010A&A...513A...7S}. 
One notable phenomenon in XDRs is that CO molecules are readily 
dissociated into C atoms, or ionized to C$^+$ or higher levels, 
which reduces CO abundance and consequently CO(2--1) intensity. 
Another effect in XDRs is that the gas temperature becomes much higher 
than in star-forming galaxies or photodissociation regions \citep[= PDRs;][]{2005A&A...436..397M}. 
We would therefore expect higher CO excitation in XDRs than in PDRs, 
which can reduce CO(2--1) line intensity by decreasing the number of 
CO molecules populating the $J = 2$ level in AGNs as compared to star-forming galaxies. 

To briefly explore the above-mentioned effect of high temperature in XDRs, 
we performed non local thermodynamic equilibrium (non-LTE) radiative transfer modelings of line intensities 
by using the RADEX code \citep{2007A&A...468..627V}. 
In order simply to grasp a qualitative trend, we performed the modelings by fixing the H$_2$ volume density ($n_{\rm H2}$) of 10$^5$ cm$^{-3}$ 
and the CO column density-to-velocity gradient ratio (this is relevant to line opacity) of $N_{\rm CO}/dV = 2.5 \times 10^{16}$ cm$^{-2}$ (km s$^{-1}$)$^{-1}$, 
which are characteristic to the CNDs of nearby galaxies \citep[e.g.,][]{2013PASJ...65..100I,2014A&A...570A..28V}. 
Throughout the modelings we adopted the cosmic microwave background temperature 
(2.73 K at $z = 0$) for the background temperature. 

Figure \ref{fig7} shows the resultant values of the CO(2--1) line intensity (brightness temperature) 
and the line opacity as a function of the gas kinetic temperature ($T_{\rm kin}$). 
Here we modeled cases of $T_{\rm kin}$ = 30, 50, 100, 200, 300, and 500 K. 
There is a trend of reducing the CO(2--1) intensity and the line opacity as increasing $T_{\rm kin}$, 
as bulk of the CO population is excited to further higher rotational $J$-levels. 
Suppose that the molecular gas temperature is typically $\lesssim 100$ K in PDRs or star-forming galaxies \citep{1997ARA&A..35..179H}, 
whereas it is much higher in XDRs \citep[at least several 100 K;][]{1996ApJ...466..561M}, 
we would expect a factor $> 2-3$ reduction in CO(2--1) intensity in AGNs than in star-forming systems 
even if the total gas mass, gas density, and CO abundance are comparable. 
Furthermore, the CO abundance is basically lower in XDRs than in PDRs, 
which further reduces the CO intensity in the former regions. 
This reduction in CO(2--1) intensity due to the high AGN luminosity of the quasar nucleus itself 
may consequently make the correlation between the CND-scale CO(2--1) luminosity 
(we rely on this to derive $M_{\rm mol}$) and the nuclear activity ambiguous, 
which would relax the tension with the observations of the lower-luminosity Seyfert galaxies. 

These XDR effects have been mainly discussed in nearby Seyfert galaxies \citep[e.g.,][]{2015ApJ...811...39I}, 
and such discussion is sparse for high redshift objects. 
However, the effects become stronger with increasing incident X-ray radiation. 
Now that we are able to probe the central CND-scale molecular gas thanks to the advent of ALMA, 
we need to carefully consider these effects to properly measure the molecular gas mass in the vicinity of luminous quasar nuclei.

\section{Summary and future prospects}\label{sec5}
We have presented high resolution CO(2--1) observations by using ALMA 
toward 4 pairs of $z < 0.1$ luminous quasars and normal star-forming galaxies, 
which are matched in redshift, $M_\star$, and SFR. 
Our prime aim is to investigate whether a systematically enhanced gas mass surface density 
is found at the circumnuclear several $\times$ 100 pc scales of the quasar sample 
as compared to the inactive comparison sample, which is predicted to be a characteristic 
initial condition for triggering AGN. 
Our conclusions are summarized as follows. 

\begin{itemize}
\item[(i)] We successfully detected the CO(2--1) emission from all quasars, 
which show diverse morphology (spiral-arm-like feature, bar-like feature, disk-like feature) in their spatial distributions. 
The bulk of the line emission we recovered 
originates from their innermost a few kpc regions. 
The line emission at the nucleus is brighter in the comparison sample 
than in the quasar sample in three of the four pairs. 
\item[(ii)] The line profile of the quasar sample is clearly broader than the matched comparison sample. 
Note that, however, we cannot provide a detailed investigation of the line profiles 
of the quasar sample (including the search for molecular outflows) given the modest S/N ratios. 
We also found that two quasars (PG1149$-$110 and likely PG1126$-$041) 
show double-horn like line profiles, which are interpreted as indications of rotating disks. 
\item[(iii)] The total molecular gas mass surface density ($\Sigma_{\rm mol}$) 
computed from the CO(2--1) line luminosity is 
accordingly higher in the comparison sample than in the quasar sample 
at the central sub-kpc regions, in three of the four pairs. 
Hence, there seems to be no systematic enhancement in $\Sigma_{\rm mol}$ in our quasars. 
This is inconsistent with our initial expectation 
that $\Sigma_{\rm mol}$ is higher in quasars than in comparison galaxies. 
\item[(iv)] We discussed four possible explanations for 
the smaller CO(2--1)-based $\Sigma_{\rm mol}$ in the quasar sample, 
i.e., AGN-driven outflows, gas-rich minor mergers, time-delay between the onsets of a starburst-phase and a quasar-phase, 
as well as X-ray-dominated region (XDR) effects on the gas chemical abundance and excitation. 
Although all of these may potentially contribute to the observed low $\Sigma_{\rm mol}$ in our quasar sample, 
we stress the importance of the XDR effects, as we started to probe 
the CND-scale molecular gas of luminous quasars thanks to the high resolution and sensitivity of ALMA. 
\end{itemize}

The time-delay we discussed in \S~4 will wash out 
a causality between circumnuclear properties and on-going quasar activities. 
Hence it is desirable to perform a test, similar to that carried out here, for AGNs 
in a much younger phase that may maintain the initial conditions to ignite the AGN. 
As for the XDR, as the actual level of the effects depends on the prevalent 
physical and chemical conditions of the CND-scale gas, 
further observations of other transition CO lines as well as other species 
sensitive to the XDR effects \citep[e.g., atomic carbon line,][]{2005A&A...436..397M,2018ApJ...867...48I} 
are required to obtain a firm conclusion 
on the importance of CND-scale gas as a fuel reservoir for SMBHs. 
In addition, for a given $\Sigma_{\rm mol}$, the gas inflow rate will increase with higher stellar mass surface density 
($\Sigma_\star$) due to stronger torques imposed \citep[e.g.,][]{2010MNRAS.407.1529H,2017MNRAS.472L.109A}. 
The high $\Sigma_\star$ is also claimed to trigger nuclear fuelings \citep{2016MNRAS.460.2360R}. 
In future we thus need high resolution stellar mass maps of both quasars and inactive galaxies 
to measure $\Sigma_\star$ and then construct a better matched sample controlled also by $\Sigma_\star$. 
This will be possible after the launch of the {\it James Webb Space Telescope (JWST)}.

\acknowledgments 
We thank the anonymous referee for her/his thorough 
reading and constructive feedback that improved this paper. 
T.I. appreciates fruitful discussion with M. Imanishi, T. Kawamuro, S. Baba, and D. Nguyen at NAOJ. 
T.I. and J.D.S. also thank L. Ho, J. Shangguan, and D. Angles-Alcazar for their help and fruitful comments. 

This paper makes use of the following ALMA data: 
ADS/JAO.ALMA\#2015.1.00872.S. 
ALMA is a partnership of ESO (representing its member states), 
NSF (USA) and NINS (Japan), together with NRC (Canada), 
MOST and ASIAA (Taiwan), and KASI (Republic of Korea), 
in cooperation with the Republic of Chile. 
The Joint ALMA Observatory is operated by ESO, AUI/NRAO and NAOJ. 

Funding for SDSS-III has been provided by the Alfred P. Sloan Foundation, the Participating Institutions, 
the National Science Foundation, and the U.S. Department of Energy Office of Science. 
The SDSS-III web site is http://www.sdss3.org/. 

SDSS-III is managed by the Astrophysical Research Consortium for the Participating Institutions 
of the SDSS-III Collaboration including the University of Arizona, the Brazilian Participation Group, 
Brookhaven National Laboratory, Carnegie Mellon University, University of Florida, 
the French Participation Group, the German Participation Group, Harvard University, 
the Instituto de Astrofisica de Canarias, the Michigan State/Notre Dame/JINA Participation Group, 
Johns Hopkins University, Lawrence Berkeley National Laboratory, Max Planck Institute for Astrophysics, 
Max Planck Institute for Extraterrestrial Physics, New Mexico State University, New York University, 
Ohio State University, Pennsylvania State University, University of Portsmouth, Princeton University, 
the Spanish Participation Group, University of Tokyo, University of Utah, Vanderbilt University, 
University of Virginia, University of Washington, and Yale University. 

W.R. is supported by the Thailand Research Fund/Office of the Higher Education 
Commission Grant Number MRG6280259 and Chulalongkorn University's CUniverse. 
T.I. is supported by the ALMA Japan Research Grant of NAOJ ALMA Project, NAOJ-ALMA-240. 
T.I. is also supported by Japan Society for the Promotion of Science (JSPS) KAKENHI Grant Number JP20K14531.

\bibliography{ref}

\appendix 
\section{Properties of one quasar-galaxy pair excluded from our test}
\restartappendixnumbering 
We excluded one quasar HE0205$-$2408 and a galaxy J0042$-$0940, which were initially paired (Table \ref{tblA1}). 
The reason for this exclusion is the mismatched SFR: 
the MPA-JHU galaxy catalog shows that the galaxy J0042$-$0940 
has $\log ({\rm SFR}/M_\odot~{\rm yr}^{-1}) \sim 1.2$ when it is measured over the galaxy-scale, 
which is comparable to the SFR estimated for the quasar HE0205$-$2408. 
However, the SFR of J0042$-$0940 measured with a 3$\arcsec$ fiber aperture 
is only $\log ({\rm SFR}/M_\odot~{\rm yr}^{-1}) \sim -2$, which is orders of magnitude 
smaller than that estimated at the galaxy-scale. 
The {\it WISE} photometry (\S~3.2) also indicates that $\log ({\rm SFR}/M_\odot~{\rm yr}^{-1}) \sim -2$ 
($W3$ band magnitude = 12.7 mag). 
Hence, we suppose that there is an unexpected error in the SFR estimation at the galaxy-scale, 
and the very small SFR measured with the fiber aperture is the correct one for this galaxy. 
Since HE0205$-$2408 and J0042$-$0940 do not compose a well-matched pair any more, 
we excluded this pair from our discussion. 

\begin{figure}
\begin{center}
\includegraphics[width=\linewidth]{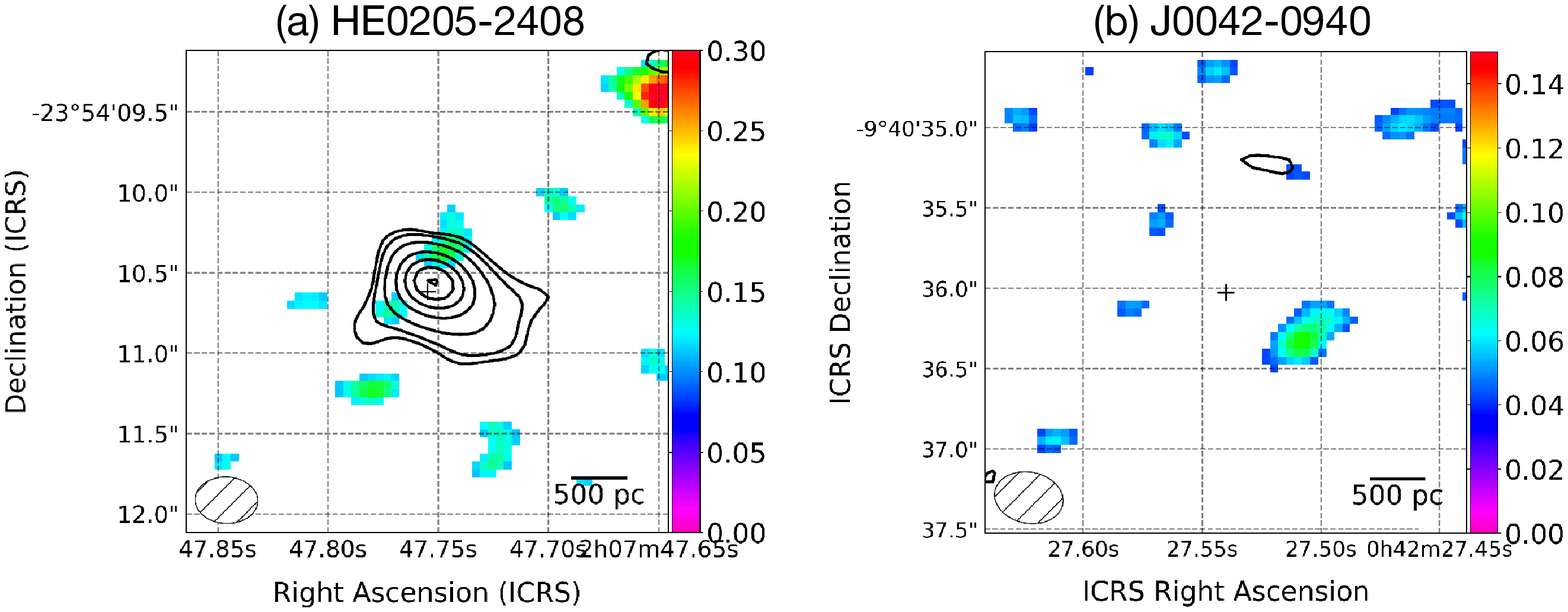}
\caption{
Velocity-integrated intensity maps of the CO(2--1) line emission (in the unit of Jy beam$^{-1}$ km s$^{-1}$) of 
the central 3$\arcsec$ region of a quasar (a) HE0205$-$2408 and a comparison galaxy (b) J0042$-$0940. 
The 1$\sigma$ sensitivity is 0.074 and 0.028 Jy beam$^{-1}$ km s$^{-1}$ for (a) and (b), respectively. 
These CO(2--1) maps are clipped at 1.5$\sigma$ level to enhance the clarity. 
The central plus signs indicate the location of the quasar nuclei or galactic centers. 
Also plotted contours indicate the underlying continuum emission distributions 
drawn at $-$3, 3, 5, 10, 20, 30, 40, and 50$\sigma$ (see text for 1$\sigma$ values). 
}
\label{fig_A1}
\end{center}
\end{figure}

We observed these objects with ALMA and analyzed the data in the same manner as described in \S~2. 
The obtained synthesized beam sizes for the CO(2--1) cube and the continuum image are, 
$0\arcsec.39 \times 0\arcsec.29$ (P.A. = 87.7$^\circ$) and $0\arcsec.37 \times 0\arcsec.29$ (P.A. = 86.6$^\circ$) for HE0205$-$2408 
and $0\arcsec.43 \times 0\arcsec.32$ (P.A. = 79.3$^\circ$) and $0\arcsec.42 \times 0\arcsec.31$ (P.A. = 79.4$^\circ$) for J0042$-$0940, respectively. 
With the natural weighting, we obtained 1$\sigma$ sensitivity of 0.30 mJy beam$^{-1}$ ($dV \sim 87$ km s$^{-1}$) 
and 0.20 mJy beam$^{-1}$ ($dV \sim 44$ km s$^{-1}$) for the CO(2--1) cubes of HE0205$-$2408 and J0042$-$0940, respectively. 
The CO(2--1) emission was undetected in both objects (Figure \ref{fig_A1}): 
the corresponding 3$\sigma$ upper limits are 0.33 Jy km s$^{-1}$ (integrated velocity width = 700 km s$^{-1}$) 
and 0.07 Jy km s$^{-1}$ (integrated velocity width = 450 km s$^{-1}$), both of which are measured at the central 700 pc region. 
On the other hand, we detected the continuum emission significantly in HE0205$-$2408 (1$\sigma$ = 52 $\mu$Jy beam$^{-1}$; max = 2.63 mJy beam$^{-1}$), 
while it is not detected in J0042$-$0940 (1$\sigma$ = 20 $\mu$Jy beam$^{-1}$). 
Given the deficit of molecular gas around the nucleus, we consider that the bulk of this continuum emission of HE0205$-$2408 to be of non-thermal origin.

\begin{table*}
\begin{center}
\caption{Properties of the quasar and the galaxy excluded from our test \label{tblA1}}
\begin{tabular}{ccccccccccc}
\tableline\tableline
\multirow{2}{*}{Name} & R.A. & Decl. & \multirow{2}{*}{$z_{\rm opt}$} & Scale & \multirow{2}{*}{log ($\frac{M_{\rm BH}}{M_\odot}$)} & \multirow{2}{*}{ log ($\frac{L_{\rm Bol}}{\rm erg/s}$) } & \multirow{2}{*}{log ($\frac{M_\star}{M_\odot}$)} & \multirow{2}{*}{log ($\frac{SFR}{M_\odot /{\rm yr}}$)}\\ 
 & (ICRS) & (ICRS) &  & (kpc/$\arcsec$) & & & & & & \\ 
\tableline
HE0205$-$2408 & 02:07:47.755 & $-$23:54:10.62 & 0.076 & 1.44 & 8.33 & 45.6 & 10.69 & 1.25 \\ 
J0042$-$0940 & 00:42:27.540 & $-$09:40:36.25 & 0.077 & 1.46 & - & - & 10.75 & $\sim -2^\dag$ \\ 
\tableline
\end{tabular}
\tablecomments{$^\dag$This is a SFR measured with a 3$\arcsec$ fiber aperture. 
As this SFR is significantly smaller than that of the paired quasar 
HE0205$-$2408, we excluded this galaxy, and consequently this pair, from the sample for our test.}
\end{center}
\end{table*}

\end{document}